\documentclass[aps,showpacs]{revtex4}
\begin{document}
\def\wick#1#2#3{\vtop{\ialign{&##\crcr
      $\displaystyle{#1}$&$\displaystyle{#2}$&
      $\displaystyle{#3}$&\crcr\noalign{\kern1pt\nointerlineskip}
      \hfil\vrule height3pt\hfil&&
      \hfil\vrule height3pt\hfil\crcr
      \noalign{\kern-.2pt\nointerlineskip}
      \hfill\hrulefill&\hrulefill&\hrulefill\hfill\crcr
      \noalign{\kern3pt}}}}
\def\bd {\begin{displaymath}}%
\def\ed {\end{displaymath}}%
\def\be {\begin{equation}}%
\def\bdc {\begin{description}}%
\def\edc {\end{description}}%
\def\ee {\end{equation}}%
\def\ba {\begin{eqnarray}}%
\def\ea {\end{eqnarray}}%
\def\bsa {\begin{subeqnarray}}%
\def\esa {\end{subeqnarray}}%
\def\bry {\begin{array}}%
\def\ery {\end{array}}%
\def\et {energy-momentum tensor}
\def\Sw {Schwinger term} 
\def\rep {reparametrisation}
\def\trf {transformation}
\def\cof {conformal} 
\def\iva {invariance}
\def\fv {field variable}
\def\fr {field redefinition}  
\def\Fr {Field redefinition}
\def\q {\quad} 
\def\qq {\qquad}
\def\nr {\nonumber}
\def\ol {\overline} 
\def\ul {\underline}
\def\ticR {\tilde{\mbox{\boldmath{$\cal R$}}}}%
\def\ti {\tilde}
\def\che {\check}
\def\ac {\acute}
\def\pt {{\partial}}
\def\Ap {{\mbox{\boldmath{{$\cal {\mbox{\LARGE$a$}}$}}}}}%
\def\Lap {{\mbox{{\Large{\bf$\ap$}}}}}%
\def\bPu {\cal {\mbox{{\Large{\bf a}}}}}
\def\Sm {{\mbox{{\LARGE$\sigma$}}}}%
\def\vp {\varphi}
\def\Bt {\mbox{\boldmath$\beta$}}
\def\cB {\mbox{\boldmath${\cal B}$}}
\def\hg {{\hat{\mbox{\boldmath{$\gamma$}}}}}
\def\ap {\alpha}
\def\zt {\zeta}
\def\cR {\mbox{\boldmath{$\cal R$}}}%
\def\cW {\mbox{\boldmath{$\cal W$}}}%
\def\bRe {\mbox{\boldmath$\Re$}}%
\def\bK {{\bf K}}%
\def\bR {{\bf R}}%
\def\bKd {{\bK_{\delta}}}%
\def\clr #1{\left( {#1}\right)}
\def\l #1{{#1}_{(l)}}
\def\bA {\mbox{\boldmath $A$}}%
\def\bW {\mbox{\boldmath $W$}}%
\def\cK {\che{\mbox{\boldmath{$\cal K$}}}}%
\def\hK {\hat{\mbox{\boldmath{$\cal K$}}}}%
\def\eps {\epsilon}
\def\ld {\lambda}
\def\lan {\langle}
\def\ran {\rangle}
\def\lbr {\lbrack}
\def\rbr {\rbrack}
\def\lbc { \lbrace }
\def\rbc {\rbrace }
\def\lrw {\longrightarrow}
\def\ci {\circ}%
\def\pr {\parallel}%
\title{Probing Hamiltonian \fr\ on the nontrivial conformal Algebra}
\author{Simon C. Lin }
\affiliation{Grid Computing Centre, Academia Sinica, 11529 Taipei, Taiwan}
\affiliation{Institute of Physics, Academia Sinica, 11529 Taipei, Taiwan}
\author{Feng-Yin Li}
\affiliation{Department of Chemistry, National Chung-Hsing University, Taichung, Taiwan }
\author{Tsong-Ming Liaw}
\email{ltming@gate.sinica.edu.tw}
\affiliation{Grid Computing Centre, Academia Sinica, 11529 Taipei, Taiwan}
\affiliation{Institute of Physics, Academia Sinica, 11529 Taipei, Taiwan}
\pacs{11.10.Ef,11.10.Kk, 11.10.Lm,11.25.Hf}

\begin{abstract}
\date{today}
The compatibility between the conformal symmetry and the closure of 
conformal algebras is discussed on the nonlinear sigma model.
The present approach, above the basis of \fr\
employed in the Hamiltonian scheme,
attempts the method of quantisation with intuitive picture.
As a general field theoretic treatment, the consistency is ensured 
by means of the interesting features which 
are observed in the historical studies for the gauge-invariant conformal symmetry. 
The identification of conformal anomaly is also shown coincident 
with the conventional one approached within the path-integral formulation. 
\end{abstract}
\maketitle
\section{introduction}
\label{one}
A generic quantisation hopefully endows with 
model-independent form-preservation rule
for operator algebras such that more physical significance of the quantised system
can be directly inferred from its classical version.
However, the simplest expectation  
that the classical and quantised operator algebras only 
deviate slightly, remains na{\"i}ve since complexity arises, occasionally.
For example, Schwinger terms \cite{tgi,swg} 
have been demonstrated for the current algebra in general.
The violation of the Jacobi identity has also been 
observed in the quark model \cite{bvr,jrw}. 
Moreover, the third cocycle is predicted 
for the quantum mechanics in the presence
of magnetic monopole \cite{gs,mc,jk0}.
While the conventional quantisation
is basically limited in an {\`a} priori renormalisation scheme,
considerable extension concerns \fr\ which is ubiquitous in the path integral approaches
\cite{ts0,as,tb,dw,gdd0,gdd,gdd1, fc0,fc1,fts,do,acny,cf,cf1,ts,mt,cl}.
Yet, concreteness can be achieved better
in the Hamiltonian scheme \cite{jw} 
where the \fr s are governed under the preservation of the canonical commutators.
Following this line,
Ref. \cite{tm} has concluded that the consistent 
quantisation must not entail arbitrary allowable configurations,
where the intriguing role of the treatment of the nonlinear sigma
model is to stress of the importance consistent quantisation over
the superficial phenomenological correspondence.
Such conclusion is sophisticate in the high energy physics \cite{jk2,chl}:
the consistent anomaly-free condition 
asserts treating the fermionic pairs equal-footingly, the so-called safe representation,
in the standard model.
Here, we continue implementing such Hamiltonian \fr\ 
to the operator algebras on the nonlinear sigma model.

In practise, 
one chooses to disregard 
the lacking of intuitive phenomenological correspondence owing
to the nonlinearity.  
Let the formulation stand on the scalar field prescription but allow it for the \fr s order by order.
At the tree level, the canonical \et\ is found symmetric and traceless, 
and the conformal invariance is sufficed \cite{jk1,ma,ccj,wl}.
As an additional feature, the form of
the canonical \et\ coincides with the
conventional stress energy tensor obtained at the flat space-time
in absence of the dilaton coupling.
Such additional feature has been 
disproved \cite{tm} at higher loop level 
since the {\em improvement} terms for achieving the symmetric and 
traceless \et\ are shown nontrivial .
However, consequently, the conserved improved \et\ implies 
the form of conformal anomaly coinciding with the two-loop Weyl anomaly 
conditions as given explicitly in, e.g., Refs. \cite{cf,cf1,ts,mt,cl}.
It has also been shown feasible when the  
torsion term and the dilaton \cite{mt,gsw} are considered.
For the former, 
the two-loop Weyl anomaly conditions are consistently coupled to the covariant torsion tensors, and 
the dependence of the residue renormalisation scheme in  anomaly conditions, stressed in Ref. \cite{mt}, is also observed.
For the latter, the dilaton part of the conformal anomaly 
can be  generated in the Hamiltonian scheme simply by means of counter-terms, even though it is poorly prescribed for
the flat space-time.
To this end, apart from the \et\, invariance properties continue to impinge on the operator algebras 
as follows. Symmetry criterion requires the conserved Noether currents,
while, in contrast, the generators for an internal symmetry transformation can 
exhibit the closed structure in absence of the symmetry \cite{chl}.
Therefore, whether the algebra is closed or not is not to encode the 
symmetry property but to comprise information peculiar to the \fv\ s in use.
Accordingly, it is not to argued model-independently, whether the closed Virasoro algebra preserves \cite{tb}
in absence of conformal anomaly for the nonlinear sigma model, and
the explicit derivation remains essential.
Our work is organised as follows.

In Section \ref{two}, the Noetherian process is reconsidered in conjunction with \fr .
It is natural to interpret the form-ambiguity of operator algebras by means of
the prescription of configuration space.
The fact that the principle of invariance permits  
such flexibility basically supports 
the historical super-potential \cite{jk1,ma,ccj} construction for the current operators.
However, it can be shown that the total divergence terms
does fully specify the corresponding ambiguity,
and the uniqueness of charge operators 
must be questioned beyond this scope.
The conclusion is also inspired by the examples based on Ref. \cite{jka,bdw}.

Section \ref{three}
revisits the previous formulation \cite{tm}
but with  more technical points relevant.
The canonical quantisation here follows the line of Ref.
\cite{jw,ds,hf,eh,hw} in the 
normal order prescription but is equipped with the convention of vacuum provided by Ref. \cite{jw}.
It is shown that the \et\  must be improved at the two loop level via the construction 
suggested in the literature \cite{jk1,ma,ccj,wl}.
The conservation of the improved \et\ then determines the conformal anomaly correctly.

In Section \ref{four}, 
the status of the conformal charges is 
clarified and the corresponding algebras 
are explicitly derived.
The generators are obtained as 
the standard Bessel-Hagen form for the conformal currents \cite{jk1,ccj,bh}.
The corresponding Fourier mode expansion
of the improved \et\ reflects the conventional holomorphic or antiholomorphic expressions
\cite{gi,za,za1,go} in the conventional two-dimensional theories.
The resultant algebras of generators, in fact, do not
reproduce the well-known Virasoro version.
The central extensions do not appear
as c-numbers in contrast to the ones appropriate for Dirac's simple conjecture \cite{go}.
The anomalous contribution is fully characterised in terms of the 
improvement terms which signify 
strong dependence on the \fv s after \fr s. 
However, the conservation of
the conformal currents remain ensured by the anomaly-free conditions.
Moreover, the basic points in Section \ref{two} appear sound  for it.

More discussions and conclusion are presented in Sec. \ref{five}.

\section{Some general features of the current and charge}\label{two}

In general, a symmetry property subject to a certain external transformation of type $T$
is prescribed by the invariance of action, say,
\begin{equation}
\delta_{T} S\ =\ \int\sum \delta_{T} \vp
\frac{\delta S}{\delta \vp},
\label{inva}
\end{equation}
where the notation $\int\sum$
denotes the functional integral over the parameter space of variation.
However, it remains flexible to determine the formulation 
in terms of distinct sets of \fv s.
In each case, the dynamical equations may end up with various fashions.
Although, such generic aspect contrasts 
with the conventional formulation for the field theory, 
which relies on a specific choice of \fv s, considerable benefits in analysis can be resulted in.
The form ambiguity for the current operators 
can be intuitively interpreted as an issue of the standard Noetherian processes 
equipped with distinct sets of \fv s.
It may turn out too restricted to decode types of ambiguity for the current operators
solely by means of the super-potential \cite{ma,jk1,ccj,wl}.
Suppose, on the contrary, it remains true,
then certain divergence term would furnish the relationship, say, 
\be
J_T^{\mu}|_B\ = \ J_T^{\mu}|_A\ +\ \pt_{\lambda}{\cal X}^{\lambda\mu},
\label{ipc}
\ee
by allowing only indistinguishable charges,
where ${\cal X}^{\lambda\mu}=-{\cal X}^{\mu\lambda}$. 
Note that $J_T^{\mu}|_B$
and $J_T^{\mu}|_A$ are the currents appropriate for the choices 
$\vp_{B}$ and $\vp_{A}$, respectively.
Equation (\ref{ipc}) cannot hold true in general among distinct sets of \fv s  
for the following argument.
Consider, for example, additional symmetry are implemented by transforming 
$\vp_{B} \rightarrow \vp_{B^U} $.
Then, the action can be as well parametrised in terms of the new 
variable $\vp_{B^U}$; 
the Noether current $J_T^{\mu}|_{B^U}$ can be explicitly deduced.
However, $J_T^{\mu}|_{B^U}$ now represents a composite transform of $T$ and the element of
$U$ which cannot be equipped with the same charge as the one for $J_T^{\mu}|_{B}$.
It turns out that
the choice of \fv s in use in fact pertains to the physical significance
for the system, an issue beyond the action itself.
Thus, the requirement on the form preservation 
of the quantised version of algebras of generators is eventually unexpected either.

One may specialise 
the flexibility discussed above to the interest for the space-time symmetry of particular types.
Above the framework of conventional formulation,
the symmetric and traceless canonical \et\ denotes 
the conformal symmetry for the system.
Such condition must be relaxed subject to the possibility 
of choosing \fv s independently.
It is sufficient to have  
the improved  \et\ of symmetric and traceless form 
by means of  
\be
{\ti T}_{\alpha\beta}\ =
T_{\alpha\beta}\ +\
W_{\alpha\beta} = T_{\alpha\beta}\ +\
\pt^{\gamma}{\cal X}_{\gamma\alpha\beta},\label{cipv}%
\ee
for a conformally invariant system \cite{jk1,wl},
where the so-called improvement term $W_{\ap\beta}$
is formally provided by the total divergence and
${\cal X}_{\gamma\ap\beta}=-{\cal X}_{\ap\gamma\beta}$.
Note that Eq. (\ref{cipv}) is a straightforward extension of Eq. (\ref{ipc}).
The situation for the conformal symmetry to be signified by the the symmetric and traceless canonical \et\
can be modified by the quantisation involving higher order;
the same condition must be imposed on the improved one appeared in Eq. (\ref{cipv}) instead. 
This is because
the consistent quantisation needs not prohibit the modification on field content by the \fr s.
Moreover, it is shown \cite{jk1} that the canonical \et\ cannot be 
a symmetric one unless the \fv\ is spinless.
Therefore,
once the the symmetric property of the canonical \et\ is not lost during quantisation,
it simply highlights that the theory does no more solely consist of scalar fields.
In order to be sufficiently generic,
the conformal current of the Bessel-Hagen form
must be considered, say,
\be
J_{\ap}=\delta_{c} x^{\beta}{\ti T}_{\ap\beta},
\label{cnc}
\ee
where $\delta_{c} x^{\beta}$ denotes 
the conformal displacements for the coordinates.
Meanwhile, the ambiguity for the operator algebras must be reduced 
after the quantisation, an effect realised by the \fr .

To this end, seemingly independent approaches of conformal currents 
performed in the two (abelian) and  more dimensional 
Yang-Mill theories \cite{jka,bdw} also shed some light to our study.
In those cases, the Noetherian process 
directly applied to the conformal transformation
fails to provide gauge-invariant conformal currents.
To furnish such structure,
the compensation by means of the gauge transformation for the vector potential
has been concluded as essential.
The issues bear interesting resemblance to our 
analysis as follows.
The deviation for the charges caused by changing \fv s is now evident.
The corresponding compensation of transform also reflects
the structure of our general argument after Eq. (\ref{ipc}).
In addition, the comparison for the case at hand with the examples couped with vector potential is 
adequate since the loss of symmetric canonical \et\ 
signifies the existence of \fv s  with nontrivial spin structure.
Moreover, the conformal algebras for the composite transformation do not close \cite{jk1,dw,jka},
rendering the corresponding finite version of transform path-dependent \cite{yang,jka}.

\section{Obtaining the conformal symmetry}
\label{three}
For the action of the nonlinear sigma model, one is referred to
\be
S\ =\ -\ \frac{T}{2}\int d\sigma d\tau\ \eta^{\ap\beta}\pt_{\ap}X^{\mu}%
\pt_{\beta}X^{\nu}G_{\mu\nu}(X), \label{action}
\ee
where $G_{\mu\nu}$ denotes the metric of the target space,
$\eta^{\ap\beta}$ is the metric of the flat world sheet with the 
signature (-1,1), 
and $T$ is often called the string tension.

The classical algebras start with the conventional choice of
\fv\ denoted by $X$ in the Hamiltonian scheme.
Formally, the typical components of the canonical \et\ can
be derived as 
\ba
T_{00}|_{X} &=& \frac{1}{2T}\Pi_{\mu}\Pi_{\nu}G^{\mu\nu}\ +\ \frac{T}{2}%
\ac X^{\mu}\ac X^{\nu}G_{\mu\nu}, \label{hcetensor1}\\
T_{01}|_{X} &=& \Pi_{\mu}\ac X^{\mu} \label{hcetensor2},%
\ea
where $\Pi$ denotes the conjugate momentum for the \fv\ $X$.
The spatial derivative is denoted 
as $\ac F(\sigma)=\pt_{\sigma} F(\sigma)$ through out this work.
Note that the Poisson brackets of the 
components given in Eqs. (\ref{hcetensor1}) and (\ref{hcetensor2}) develop a
simple closed structure for the classical algebras.
Moreover, 
the \et\ is traceless and symmetric in both its indices, namely,
\ba 
T^{\ap}_{{\ }\ap}|_{X} &\equiv & T_{00}|_{X}\ -\ T_{11}|_{X}\equiv 0, 
\label{ctrace}\\ 
T_{01}|_{X} &\equiv & T_{10}|_{X}, \label{csym}
\ea
which automatically fulfil the conformal symmetry conditions at
the classical level.

For quantisation, 
the normal ordering is introduced in 
order to subtract the tadpoles.
Hence, the operator products
exhibit the standard decomposition 
\be
A(\sigma)B({\ti \sigma})\ =\ :A(\sigma)B({\ti \sigma}):\
+\ \wick{A}{(\sigma)}{B}({\ti \sigma}), \label{contract}
\ee
where the normal ordered product is delimited
by a pair of double dots and the Wick-contraction is underlined.
The Fourier expansion is employed, say, 
\be
O^{m}\ =\ \frac{1}{2}\int_{0}^{\pi} d\sigma e^{-2im\sigma}O(\sigma)\ =\ -\frac{i
}{4}\oint
d\xi \xi^{-m-1} O(\sigma)\ = \int_{m}O(\sigma) \label{mode}
\ee
with $\xi=e^{2i\sigma}$ and $\ti\xi=e^{2i\ti\sigma}$,
such that any equal-time commutator is converted into contracted form
as follows,
\be
\left\lbr A^{m},\ B^{n} \right\rbr\ 
= -\ \frac{1}{16}\oint d\xi \oint d \ti\xi \xi^{-m-1}%
 \ti\xi^{-n-1} \wick{A}{(\xi)\ }{B}(\ti\xi),\ m,n,\in Z,
\label{qwick}
\ee
where $A$ and $B$ are considered as functionals of the fundamental
conjugate pair $\Pi$ and $X$.
Note that the upper indices $m$ and $n$ 
in Eqs. (\ref{mode}) and (\ref{qwick}) simply stand
for the corresponding Fourier mode and 
should not be confused with the power of the objects.
The contracted term on the R.H.S. of Eq. (\ref{qwick})
can be fully determined,
once the fundamental contraction for $\Pi$ and $X$
is fixed by a proper order prescription.
Upon renormalising the vacuum with the convention of 
Ref. \cite{jw}, the fundamental contraction reads
\be
\wick{\Pi}{_{\mu}(\sigma)}{X}^{\nu}(\ti\sigma) = -\ \frac{i}{\pi}\ \delta_{
\mu}^{\nu}\ \frac{\ti\xi}{\ti\xi-\xi} \label{wickc}
\ee
subject to the condition $|\xi|<|\ti\xi|$.
This renders the quantised Hamiltonian at the tree level 
formally unaltered in contrast to the classical one \cite{jw}.

The \fr\ can be implemented by writing
the generic canonical \et\ as
\be
T_{\ap\beta}\ =\ T_{\ap\beta}|_{ X}\ +\  T^{II}_{\ap\beta}\
+\  T^{III}_{\ap\beta}\ +\ \cdots,
\label{ect}
\ee
where $T_{\ap\beta}|_{X}$'s  $(\alpha=0,1)$ 
corresponds to the canonical
\et\ obtained from the tree level. 
The counterterms are known to be 
\ba
\clr{T_{00}^{II}}^{m} &=& \frac{-i(m+1)}{2\pi T} \int_{m}\ \Pi \Gamma,\label{ec1}\\
\clr{T_{00}^{III}}^{m}&=& \frac{1}{4\pi^2 T}\int_{m}\ \Pi G
\clr{\cR +\Lap}{\ac X}\
- \frac{m(m+1)(k+1)}{12\pi^2 T}\int_{m}\ tr\clr{G\cR +G\Lap},\label{ec2} \\
\clr{T_{00}^{IV}}^{m} &=& \frac{i(m+1)}{8\pi^2 T}\int_{m} Q{\ac X},
\label{ec3}\\
\clr{T_{01}^{II}}^{m}&=& \frac{i(m+1)}{2\pi}\int_{m}\ {\bar \Gamma}{\ac X},
\label{ec4}
\ea
while $T_{01}^{III}$ and $T_{01}^{IV}$ remain trivial.
The explicit manipulation 
picks up the zeroth order of antisymmetric coupling
from of the algebras of the canonical \et\ \cite{tm} and shows 
\ba
\lbr\ \clr{T_{00}}^{m},\ \clr{T_{00}}^{n}\ \rbr &=& (n-m)\clr{T_{01}}^{n+m}\ +\
\frac{D}{6}(m^3-m)\delta_{m+n,0}\nr\\
&+& \frac{(n-m)}{2 \pi}\int _{n+m}{\ac X}\cR {\ac X}\ +\ 
\frac{(n-m)}{8 \pi^2 T^2}\int_{n+m} \Pi G{\cal U}{\ac X}\nr\\
&+& \frac{i(n-m)(n+m+1)(k-2)}{24\pi^2 T^2}\int_{n+m}\Pi G\pt\ tr\clr{G\cR } \nr\\
&+& \frac{(n^3-n)-(m^3-m)}{16 \pi^3 T^2}{\cal I}^{n+m}\ +\ 
\frac{(n-1)(n+1)(m+1)}{16 \pi^3 T^2}{\cal S}^{n+m}\nr\\
&+& O(5)+\cdots,\label{oag1}\\ 
\lbr\ \clr{T_{00}}^{m},\ \clr{T_{01}}^{n}\ \rbr  &=& (n-m)\clr{T_{00}}^{n+m}
\nr\\
&+& \frac{m(m+1)(n+1)(2-k)-m(m^2-1)k}{12\pi^2 T} \int_{n+m}\ tr\clr{G\cR }\nr\\
&+& O(5)+\cdots,\label{oag2}\\
\lbr\ \clr{T_{01}}^{m},\ \clr{T_{01}}^{n}\ \rbr &=& (n-m)\clr{T_{01}}^{n+m}\ +\
\frac{D}{6}(m^3-m)\delta_{m+n,0}\label{oag3}
\ea
with 
\ba
{\cal U}_{\mu\nu} &=& 
\left \lbc\ \pt_{\mu}G^{\rho\tau}\pt_{\rho}\left( G(\Lap +\cR) \right)_{\ \mu}%
^{\epsilon} - \pt_{\rho}G^{\epsilon\tau}\pt_{\tau}%
 \left( G(\Lap +\cR) \right)_{\ \mu}^{\rho} + \pt_{\rho}G^{\epsilon\tau}%
\pt_{\mu}\left( G(\Lap +\cR) \right)_{\ \tau}^{\rho}
\right \rbc\ G_{\epsilon\nu}\nr\\
&-& \left \lbc\ \pt_{\mu}\pt_{\lambda}G^{\rho \tau}\pt_{\rho}\pt_{\tau}%
G^{\lambda\epsilon} + \pt_{\rho}G^{\epsilon\lambda}\pt_{\mu}\pt_{\lambda}%
\Gamma^{\rho}\right \rbc\ G_{\epsilon\nu}\nr\\
&+& \frac{1}{2}\ \left \lbc\ G^{\rho\tau}\pt_{\rho}\pt_{\tau}
\left( G(\Lap +\cR) \right)_{\ \mu}^{\epsilon} - \pt_{\mu}G^{\rho\tau}%
\pt_{\rho}\pt_{\tau}\Gamma^{\epsilon} + \Gamma^{\rho}\pt_{\rho}\left( G(
\Lap +\cR) \right)_{\ \mu}^{\epsilon}\right.\nr\\
&-& \left. \pt_{\rho} \Gamma^{\epsilon}\left( G(\Lap +\cR) 
\right)_{\ \mu}^{\rho} + \pt_{\mu}\Gamma^{\rho}\left( G(\Lap +\cR) 
\right)_{\ \rho}^{\epsilon}\ \right\rbc\ G_{\epsilon\nu}\nr\\ 
&+& \frac{1}{2}\left\lbc\ 
\left( G(\Lap+\cR)G(\Lap+\cR)G \right)_{\mu\nu} +
 \left( \pt_{\nu} Q_{\mu} -\pt_{\mu} Q_{\nu} \right)\ \right\rbc,\label{Uep}\\ 
{\cal I} &=& 
\frac{2}{3}\ \pt_{\rho}\left( G(\Lap +\cR) \right)_{\ \lambda}%
^{\mu}\pt_{\mu}G^{\lambda\rho} - \frac{1}{6}\left\lbc\pt_{\lambda}\pt_{\rho}%
G^{\mu\nu}\pt_{\mu}\pt_{\nu} G^{\lambda\rho} 
+ tr\left( G(\Lap+\cR)G(\Lap+\cR) \right)\right\rbc,\label{Iep}\\
{\cal S} &=& 
\frac{1}{3} \left( \Gamma^{\mu}\pt_{\mu} + G^{\mu\nu}\pt_{\mu}\pt_{\nu}\right)
\left( \left(k+1\right)tr(G\cR )+ tr\left(G\Lap\right)\right) 
- \left( G(\Lap +\cR) \right)_{\ \nu}^{\mu}\pt_{\mu}\Gamma^{\nu}%
\nr\\
&+& \pt_{\lambda}\pt_{\rho}\Gamma^{\mu}\pt_{\mu}
G^{\lambda\rho}+\pt_{\mu}\Gamma^{\nu}\pt_{\nu}\Gamma^{\mu}%
+ G^{\lambda\rho}\pt_{\lambda}Q_{\rho} +\Gamma^{\mu}Q_{\mu},\label{Sep}
\ea
where
\ba
\Lap_{\mu\nu} &=& \frac{1}{4}\pt_{\mu}G^{\rho\tau}\pt_{\nu}G_{\rho\tau}-
\frac{1}{2}\left( \pt_{\mu}G^{\rho\tau}\pt_{\tau}G^{\rho\nu}+\pt_{\nu}%
G^{\rho\tau}\pt_{\tau}G_{\rho\mu}\right)\nr\\
&+& \frac{1}{2}\left(G^{\lambda\tau}G^{\epsilon\rho} -
G^{\rho\tau}G^{\epsilon\lambda}\right)\pt_{\rho}G_{\nu\lambda}%
\pt_{\tau}G_{\mu\epsilon},\label{Lapep}\\
Q_{\lambda} &=& -\Gamma_{\mu}\left( G(\Lap +\cR) \right)^{\mu}_{\ \lambda}%
-G^{\rho\mu}\pt_{\rho}\left( G(\Lap +\cR) \right)^{\mu}_{\ \lambda}+
\pt_{\lambda}{\bar \Gamma}_{\mu}\pt_{\rho}G^{\mu\rho},
\label{Qep}
\ea
where $\cR_{\mu\nu} $ is the Ricci tensor and various contracted connections 
have been employed as
\bd
\Gamma^{\ld}\ =\ \Gamma_{\mu\nu}^{\ \ \ld}G^{\mu\nu},\qq  
\bar \Gamma_{\mu}\ =\ \Gamma^{\ \rho}_{ \rho\ \mu}
\ed
and the covariant derivative is defined as 
$D_{\mu}V_{\nu}= \pt_{\mu}V_{\nu}+\Gamma_{\mu\nu}^{\ \ \ld}V_{
\ld}$.
Note that the expressions of Eqs. (\ref{ec1} - \ref{Qep}) 
have been compacted by letting the neighbouring indices contracted.
Hence, objects appearing in the parenthesis of $tr(\cdots)$ are contracted cyclically.
In addition, the spatial derivatives are abbreviated in form
of $\ac F(\sigma)=\pt_{\sigma} F(\sigma)$ through out this work.
In Eqs. (\ref{oag1} -\ref{oag3}), corrections to higher orders are denoted with $O(N)$,
corresponding to terms of $N$ derivatives with respect to the \fv s.
                                                                                                                             
The expressions of Eqs. (\ref{Uep} - \ref{Qep})
have also been partially shown in Ref. \cite{jw}.
The basic observation is that 
the matrix ${\cal U}_{\mu\nu}$, containing antisymmetric part,
can not be made covariant simply by extending counterterms with higher order for the
\fr .
This problem is remedied by the consideration 
of Eq. (\ref{cipv}) which renders the conformal symmetry
signified by the improved \et\ instead.
Resembling the criterion appropriate
for the general choice of \fv , 
the conditions are, via the obtainable conserved improvement terms, 
\ba 
\ti T^{\ap}_{\ \ap} &=& 0,\label{qtrc}\\
\ti T_{\ap\beta} &=& \ti T_{\beta\ap},\label{qsymm} \\
\pt^{\ap}\ti T_{\ap\beta} &=& 0\label{qconserv}.
\ea

Accordingly, Eqs. (\ref{oag1} -\ref{oag3}) are identified
as 
\ba
\lbr\ (T_{00})^{m},(T_{00})^{n}\ \rbr &=& (n-m)(T_{01})^{n+m}\ 
-(n-m)\left( \ti T_{01}\ -\ \ti T_{10} \right)^{n+m}\ +\ c_{1}(n,m)\nr\\*
 &+& (n-m)(W_{01}\ -\ W_{10})^{n+m},\label{gal1}\\
\lbr\ (T_{01})^{m},(T_{00})^{n}\ \rbr &=& (n-m)(T_{00})^{n+m}\ 
+\ m\ \left( \ti T^{\ap}_{\ap}\right)^{n+m}\ +\ c_{2}(n,m)\nr\\*
 &-& m\ (W_{11}\ -\ W_{00})^{n+m},\label{gal2}\\ 
\lbr \ (T_{01})^{m},(T_{01})^{n}\ \rbr &=& (n-m)(T_{01})^{n+m}\ +\ c_{3
}(n,m), \label{gal3}   
\ea
where the functions $c_{i}(n,m)$'s
are restraint by the translational invariance.
Although this  obviously deviates from the closure version
assumed by the classical formulation subject to the conventional
\fv\ $X$, the symmetry conditions can be achieved as well, namely, 
\ba
 \lbr \ (T_{00})^{0},(\ti T_{00})^{n}\ \rbr &=& n\ (\ti T_{01})^{n}\ 
 -\ n\ (\ti T_{01}\ -\ti T_{10})^{n},\label{ane1}\\
 \lbr \ (\ti T_{01})^{m},(T_{00})^{0}\ \rbr &=& -m\ (\ti T_{00})^{m}\ -\ 
 m\ (\ti T^{\ap}_{\ap})^{m},\label{ane2} %
\ea
which indicate that the divergence of the improved current 
is up to the quantities 
$n(\ti T_{01}-\ti T_{10})^{n}$ and $m(\ti T^{\ap}_{\ap})^{m}$.
To be concrete, once the symmetry is recovered, 
$\ti T_{\ap\beta}$ must conserve and 
its symmetric and traceless property can be consistently achieved.

The peculiar structure of the improvement term $W_{\alpha\beta}$
in two dimensions \cite{wl,tm}
also reflects the conventional construction as shown in Eq. (\ref{cipv}).
Basically, the corresponding construction is 
\ba
\clr{{\bar W}_{00}}^h &=& \frac{ih}{8\pi^2 T}\int_{h}
             \left(\ {\ac X}Y+\frac{1}{T}\Pi GY \right)
             \ -\ \frac{1}{8\pi^2 T}\int_{h}{\bar F} \nr\\
         &+& \frac{h^2-1}{24\pi^3 T^2}\int_{h}\ tr\left( G({\cal E}_{1}+
             {\cal E}_{3})\right),\label{tiprv00}\\ 
\clr{{\bar W}_{01}}^h &=& \frac{ih}{8\pi^2 T}\int_{h}
             \left(\ {\ac X}Y+\frac{1}{T}\Pi GY \right)
             \ -\ \frac{1}{8\pi^2 T}\int_{h} F \nr\\
         &+& \frac{h^2-1}{24\pi^3 T^2}\int_{h}\ tr\left( G(
             {\bar{\cal E}}_{1}+ {\bar{\cal E}}_{3})\right) 
             + \frac{h(h+1)}{16\pi^3 T^2}\int_{h}{\cal S}\label{tiprv01}
\ea
with
\ba
{\bar F} &=& \frac{-1}{8\pi^2 T} \left(\ {\ac X}{\bar
{\cal E}}_{1}{\ac X}+ \frac{1}{T}\Pi G{\bar {\cal E}}_{2}{\ac X}+
\frac{1}{T^2}\Pi G{\bar {\cal E}}_{3} G\Pi \right), \label{bF01}\\
{F} &=& \frac{-1}{8\pi^2 T} \left(\ {\ac X}{\cal
E}_{1}{\ac X}+ \frac{1}{T}\Pi G{\cal E}_{2}{\ac X}+ \frac{1}{T^2}\Pi G
{\cal E}_{3}G\Pi \right)\label{F01},
\ea
where
$Y$, ${\bar {\cal E}}_{i}$'s and ${\cal E}_{i}$'s are considered as
functional of the \fv\ $X$ of proper ranks, $Y$ is of $O(3)$ and
${\cal E}_{i}$'s, ${\cal E}_{i}$ are of $O(4)$.
The other two components of the improvement term are similarly determined 
by conservation laws as
\ba
\clr{{\bar W}_{10}}^h &=& \frac{1}{h}\left\lbr\ \clr{T_{00}}^{0}, \ 
\clr{{\bar W}_{00}}^h\ \right\rbr\nr\\ 
&=& \frac{ih}{8\pi^2 T}\int_{h}\left(\frac{1}{T}\Pi GY+\ac X Y \right)
+ \frac{h^2-1}{24\pi^3T^2}\int_{h} tr\left(G\left(
{\bar {\cal E}}_{1}+ {\bar {\cal E}}_{3} \right)\right)\nr\\
&+& \frac{1}{8\pi^2 T^2}\int_{h} \Pi G \left( \frac{1}{2}\left(DY-(DY)^T\right)
+\left(-2{\bar \ap}_{1}{\bar {\cal E}}_{1}-2{\bar \ap}_{2}{\bar {\cal E}}_{3} 
\right) \right)\ac X\nr\\
&+&\frac{1}{8\pi^2 T}\int_{h} \left( \frac{1}{T^2}\Pi G\left(
\frac{1}{2}DY-{\bar \ap}_{3}{\bar {\cal E}}_{2}\right)G\Pi
+\ac X\left(-\frac{1}{2}DY-{\bar \ap}_{4}{\bar {\cal E}}_{2} 
\right)\ac X\right),\label{tiprv10}\\
\clr{{\bar W}_{11}}^h &=& \frac{1}{h}\left\lbr\ \clr{T_{00}}^{0}, \ 
\clr{{\bar W}_{01}}^h\ \right\rbr\nr\\
&=& \frac{ih}{8\pi^2 T}\int_{h}\left(\frac{1}{T}\Pi GY+\ac X Y \right) 
+ \frac{h^2-1}{24\pi^3T^2}\int_{h} tr\left(G\left({\cal E}_{1}+
{\cal E}_{3} \right)\right)\nr\\ 
&+& \frac{1}{8\pi^2 T^2}\int_{h} \Pi G \left( \frac{1}{2}\left(DY-(DY)^T\right)
+\left(-2\ap_{1}{\cal E}_{1}-2\ap_{2}{\cal E}_{3} \right)
\right)\ac X\nr\\ 
&+&\frac{1}{8\pi^2 T}\int_{h} \left( \frac{1}{T^2}\Pi G\left(
\frac{1}{2}DY-\ap_{3}{\cal E}_{2}\right)G\Pi 
+\ac X\left(-\frac{1}{2}DY-\ap_{4}{\cal E}_{2} \right)\ac X\right)\label{tiprv11}.
\ea
Note that, in above, the undetermined constants $\ap_{i}$'s and $\bar\ap_{i}$'s
($i=1..4$) respond to the choice of the $O(5)$ term 
for the single $\lbr (T_{00})^m, O^n\rbr$ type of contractions
in the sense that
\ba
\int_{m}\int_{n}:\wick{\Pi} {G\Pi:\ :\ac X {\bar{\cal E}_{1}} } {\ac X}:
&=& 
\frac{-n\pi}{2}\int_{m+n}:\Pi G{\bar{\cal E}_{1}}\ac X:+
\frac{-i\pi}{4}\int_{m+n}:\Pi G\frac{d}{d\sigma}({\bar{\cal E}_{1}}\ac X):\nr\\
&=&
\frac{m\pi}{2}\int_{m+n}:\Pi G{\bar{\cal E}_{1}}\ac X:+
\frac{i\pi}{4}\int_{m+n}:\frac{d}{d\sigma}(\Pi G){\bar{\cal E}_{1}}\ac X:\nr\\
&=&
\frac{(\gamma(-n)+(1-\gamma)m)\pi}{2}\int_{m+n}:\Pi G{\bar{\cal E}_{1}}
\ac X:\nr\\
&+&\frac{-i\gamma\pi}{4}\int_{m+n}
:\Pi G\frac{d}{d\sigma}({\bar{\cal E}_{1}}\ac X):
+\frac{i(1-\gamma)\pi}{4}\int_{m+n}:\frac{d}{d\sigma}(\Pi G)
{\bar{\cal E}_{1}}\ac X:,\label{amb} 
\ea
where the first two equations deviate from each other 
only by a surface term
and the generic form of the third equation can be linear combination of above two equations, weighted 
by the constant $\gamma$.

In addition, modifications on the zeroth mode should be included 
such that the status for Hamiltonian and the translation generator on the space direction 
can remain.
This amounts to
\ba
\clr{W_{00}}^h&=&
=\clr{{\bar W}_{00}}^h(1-\delta_{h,0}), \label{fipv1}\\
\clr{W_{01}}^h&=&
=\clr{{\bar W}_{01}}^h(1-\delta_{h,0}), \label{fipv2}\\
\clr{W_{10}}^h&=&
=\clr{{\bar W}_{10}}^h-\clr{{\bar W}_{01}}^h\delta_{h,0},\label{fipv3}\\
\clr{W_{11}}^h&=&
=\clr{{\bar W}_{11}}^h-\clr{{\bar W}_{00}}^h\delta_{h,0}.\label{fipv4}
\ea
Noteworthy is that
\be
\left(\ \lbr\ \clr{T_{00}}^0,\ \clr{{\bar W}_{0\beta}}^h\ \rbr-h\clr{{\bar W}_{1\beta}}^h\ \right)
=\left(\left\lbr\ \clr{T_{00}}^0,\ \left( W_{0\beta} \right)^h\ \right\rbr
-h\left( W_{1\beta} \right)^h \right),\ \forall h,
\ee
leaving the derivations for Eqs. (\ref{tiprv10}) and (\ref{tiprv11}) unaltered. 
Moreover,
since
\ba
{W}_{01}-{W}_{10} &=& {\bar W}_{01}-{\bar W}_{10},\\
{W}_{00}-{W}_{11} &=& {\bar W}_{00}-{\bar W}_{11},
\ea
considerable simplifications can be achieved by the observation
that Eqs. (\ref{gal1}) and (\ref{gal2}) only depend on the difference between
the improvement terms.

Subsequently, the improvement term 
is proposed such that the antisymmetric part of 
$\Pi G{\cal U}{\ac X}$ coupling in Eq. (\ref{oag1}) is cancelled out 
via further ajouting the counterterms, 
$T_{\alpha\beta}^{V}$ to the expressions appropriate for  Eqs. (\ref{ec1} - \ref{ec4}). 
Hence, consistency equations relating
the known quantities in Eqs. (\ref{oag1} - \ref{oag3}) to the
improvement terms  are determined as
\ba
\clr{{\bar {\cal E}}_{1}}_{(\mu\nu)}-\frac{1}{2}D_{(\mu}Y_{\nu)}-
\ap_{4}\clr{{\cal E}_{2}}_{(\mu\nu)}&=&0, \label{con1}\\ 
\clr{{\bar {\cal E}}_{2}}_{(\mu\nu)}-2\ap_{1}\clr{{\cal E}_{1}}_{(\mu\nu)}-
2\ap_{2}\clr{{\cal E}_{3}}_{(\mu\nu)}&=&0,\label{con2}\\
\clr{{\bar {\cal E}}_{3}}_{(\mu\nu)}-\ap_{3}\clr{{\cal E}_{2}}_{(\mu\nu)}+
\frac{1}{2}D_{(\mu}Y_{\nu)}&=&0,\label{con3}\\
\clr{{\cal E}_{1}}_{(\mu\nu)}-D_{(\mu}Y_{\nu)}-
\clr{{\bar\ap}_{4}-{\bar\ap}_{3}}{\bar {\cal E}}_{2}-
\clr{{\cal E}_{3}}_{(\mu\nu)}&=&
{\cal K}_{\mu\nu},\label{con7}\\
\clr{{\bar {\cal E}}_{2}}_{[\mu\nu]}-\frac{1}{2}D_{[\mu}Y_{\nu]}&=&0,
\label{con4}\\
{\cal U}_{[\mu\nu]}+\clr{{\cal E}_{2}}_{[\mu\nu]}
-\frac{1}{2}D_{[\mu}Y_{\nu]}&=&0,\label{con5}\\
tr\left( G\left({\cal E}_{2}-
2{\bar \ap}_{1}{\bar {\cal E}}_{1}-2{\bar \ap}_{2}{\bar {\cal E}}_{3}
\right)\right) &=& 3\clr{{\cal S}-{\cal I}}-tr\clr{G{\cal U}},\label{con6}\\
\frac{1}{2}DY+tr\left( G\left({\cal E}_{3}-
{\bar \ap}_{3}{\bar {\cal E}}_{2}\right)\right) &=& 
tr\left( G{\cal K} \right),\label{con8}   
\ea
where ${\cal K}_{\mu\nu}$ denotes
a certain covariant functional of $O(4)$ and the notation 
of covariant derivative $D_{\mu}Y_{\nu}$ are employed for compacting.
In addition, 
the choice of $k=2$ in Eq. (\ref{oag1} - \ref{Sep}) 
turns out to be the scheme of \fr\ appropriate for the Principle of Least Anomaly \cite{tm}.
Hence,
\be
\clr{T_{00}^V}^{h} = \frac{-1}{16\pi^2 T}\int_{h}{\ac X}
\left( {\cal U} + {\cal E}_{2}- 2{\bar \ap}_{1}{\bar {\cal E}}_{1}-
2{\bar \ap}_{2}{\bar 
{\cal E}}_{3}\right){\ac X} + \frac{-1}{8\pi^2 T^2}\int_{h}\Pi G 
\left( {\cal E}_{3} - {\bar\ap}_{3}{\bar {\cal E}}_{2}+\frac{1}{2}\left(
DY+(DY)^T\right)
 \right){\ac X},  \label{t5} 
\ee
while no further counterterms for $T_{01}$ are needed.
Consequently,
\ba
\lbr\ \clr{T_{00}}^{m},\ \clr{T_{00}}^{n}\ \rbr &=& 
(n-m)(T_{01})^{n+m}\ +\
\frac{D}{6}(m^3-m)\delta_{m+n,0}\nr\\
&+& \frac{(n-m)}{2 \pi}\int _{n+m}{\ac X}\left( \cR +\frac{1}{4\pi T}{\cal K}%
\right){\ac X}\nr\\ 
 &+& (n-m)({\bar W}_{01}\ -\ {\bar W}_{10})^{n+m}\nr\\
 &+& O(5) +\cdots,\label{nag1}\\
\lbr\ \clr{T_{00}}^{m},\ \clr{T_{01}}^{n}\ \rbr  &=& (n-m)\clr{T_{00}}^{n+m} 
- \frac{m(m^2-1)}{6\pi^2 T} \int_{n+m}\ tr\clr{G\cR} \nr\\
 &-& m \clr{{\bar W}_{11}\ -\ {\bar W}_{00}}^{n+m}\nr\\
 &+& O(5)+\cdots,\label{nag2}\\
\lbr\ \clr{T_{01}}^{m},\ \clr{T_{01}}^{n}\ \rbr &=& (n-m)\clr{T_{01}}^{n+m}\ +\
\frac{D}{6}(m^3-m)\delta_{m+n,0}.\label{nag3}
\ea   

In above, the consistent identifications
for Eqs. (\ref{ane1}) and (\ref{ane2})
provide the anomaly conditions, up to $O(5)$, as
\ba
n \clr{\ti T_{01}\ -\ \ti T_{10}}^{n}\ &=& \frac{n}{2\pi}\int_{n}%
\ac X\left( \cR + \frac{1}{4\pi T}{\cal K} \right)\ac X, \\
m \clr{{\ti T}^{\ap}_{\ap}}^{m} &=& \frac{-m(m^{2}-1)}{6\pi^{2}T}%
\int_{m}  tr\left(G \clr{\cR + \frac{1}{4\pi T}{\cal K}} \right) ,\label{anmc2}
\ea
which suggest that the conformal symmetry can be recovered
when the condition 
\be 
\cR_{\mu\nu} + \frac{1}{4\pi T}{\cal K}_{\mu\nu}+\ 
O(6)\ + \cdots\ =\ 0
\label{fanmc}
\ee
is fulfilled.
However, the expression can be cast more limited
by virtue of the recursive use of its consistency conditions   
and the secondary constraints \cite{dirac} thereof.
${\cal K}_{\mu\nu}$ only
appears as a linear combinations of covariant quantities of $O(4)$, say,
\be
\cK_{\mu\nu}=\sum_{i}\ap_{i}\cK^{(i)}_{\mu\nu},\label{klp}
\ee
in which some of freedom can be further subtracted
out by means of the consistent conditions of Eq. (\ref{fanmc}).
The corresponding conditions 
relate $O(4)$ to $O(6)$ appears as direct consequence 
of Eq. (\ref{fanmc}) and  are given, for instance, as
\ba
\cR_{\mu\lambda}\cR^{\lambda}_{\ \nu}+\frac{1}{4\pi T}\cR_{\mu\lambda}
\cK^{\lambda}_{\ \nu}+\cdots &=& 0,\\
\bRe_{\mu\ld\rho\nu}\cR^{\ld\rho}+\frac{1}{4\pi T}\bRe_{\mu\ld\rho\nu}{\cal K}^{\ld\rho}+\cdots&=&0,\\
D^2\cR_{\mu\nu}+\frac{1}{4\pi T}D^2{\cal K}_{\mu\nu}+\cdots&=&0,\\
D^{\ld}D_{\mu}\cR_{\ld\nu}+\frac{1}{4\pi T}D^{\ld}D_{\mu}{\cal K}_{\ld\nu}+\cdots&=&0,
\ea
where
the first two are obtained via direct contractions and 
the latter two are derived from the secondary constraints \cite{dirac}. 
While no other relevant quantities essentially enter,
except for the dependence on the conventional Riemann $\bRe_{\mu\ld\rho\tau}$,
one can achieve,
\be
\cR_{\mu\nu}\ +\ \frac{\ap}{4\pi T}\bRe_{\mu}^{\ \ld\rho\tau}\bRe_{\nu\ld\rho
\tau}\ + O(6)+\cdots =\ 0, \label{tfanmc}
\ee
where $\ap$ remains some undetermined constant.
Noteworthy is about 
the object $D^{\lambda}D^{\rho}\bRe_{\mu\lambda\rho\nu}$
that could have contributed to Eq. (\ref{fanmc})
via Eq. (\ref{klp}).
However, this decomposed, identically, as 
\be
D^{\lambda}D^{\rho}\bRe_{\mu\lambda\rho\nu}=
D^2\cR_{\mu\nu}-D^{\ld}D_{\mu}\cR_{\ld\nu},
\ee
after employing the Bianchi identity.
In addition, the missing of the $O(5)$ term is concluded by 
the fact that covariant objects of $O(5)$ do not appear as matrices.


Arbitrariness left in
the counterterms of \fr\ 
is referred to as permissible schemes for the quantisation. 
The corresponding ambiguity is reduced by means of 
the minimal setting of the anomaly \cite{tm}
which amounts to possibly weakening the anomaly conditions.
Meanwhile, this ensures the faithfulness of 
the resultant essential symmetry requirement. 
Subject to this principle,  schemes other than $k=2$
may be adopted in Eqs. (\ref{oag1} - \ref{oag3}),
instead.
As the consequence, the anomaly terms will be displaced, 
moving, to and fro, between ${\ti T}_{\ap}^{\ \ap}$ and ${\ti T}_{01}-{\ti T}_{10}$.
However, the resultant condition must remain unaltered.
The situation also resembles
the treatment of Weyl anomaly pedagogically reviewed in  Ref. \cite{cl}
where all the anomalies can be shifted to the trace part of \et .


\section{The Conformal Algebras}
\label{four}
After clarifying the status of \et\, we  
turn to the direct study of the structure of generators of transformations.
For the two dimensions,
the conformal generators 
are known to be provided in terms of holomorphic and 
antiholomorphic expressions \cite{gi,za,za1}.
According to the convention of \cite{gi}, 
the forms of conformal generators $L_{m}$ are given as
\ba
L_{m}&=& \oint d\zeta \zeta^{m-1} {\cal T}(\zeta)=\int dx^{1} e^{mz}T(z),
\label{gil1}\\
{\bar L}_{m}&=& \oint d{\bar \zeta} {\bar \zeta}^{m-1}{\bar {\cal T}}
({\bar \zeta})=\int dx^{1} e^{m{\bar z}}{\bar T}({\bar z}),\label{gil2}
\ea
where
\ba
z&=& x^1+ ix^0, \label{z1}\\ 
{\bar z} &=& x^1- ix^0,\label{z2}\\
\zeta&=& \exp z,\\
{\bar \zeta}&=& \exp {\bar z}.
\ea 
and the contour integrations in Eqs. (\ref{gil1}) and  (\ref{gil2}) make sense 
of the radial quantisation discussed in the context of conformal field theory. 

On the other hand, the treatment of Eqs. (\ref{fipv1} - \ref{fipv4})
plays the crucial role of conservation relation 
for the conformal currents, namely,
\ba
\left\lbr\ \clr{{\ti T}_{00}}^0,\ \clr{{\ti T}_{00}}^n\ \right \rbr&=& 
\left\lbr\  \clr{T_{00}}^0,\ \clr{{\ti T}_{00}}^n\ \right \rbr\ =\
 n\ \clr{{\ti T}_{01}}^{n},\label{csv1}\\
\left\lbr\ \clr{{\ti T}_{01}}^0,\ \clr{{\ti T}_{00}}^n\ \right \rbr &=&
\left\lbr\  \clr{T_{01}}^0,\ \clr{{\ti T}_{00}}^n\ \right \rbr\ =\
 n\ \clr{{\ti T}_{00}}^{n},\label{csv2}\\
\left\lbr\ \clr{{\ti T}_{01}}^m,\ \clr{{\ti T}_{00}}^0\ \right \rbr &=&
\left\lbr\ \clr{{\ti T}_{01}}^m,\  \clr{T_{00}}^0\ \right \rbr\ =\
-m\clr{{\ti T}_{00}}^{m},\label{csv3}\\ 
\left\lbr\ \clr{{\ti T}_{01}}^0,\ \clr{{\ti T}_{01}}^m\ \right\rbr &=& 
\left\lbr\ \clr{T_{01}}^m,\ \clr{{\ti T}_{01}}^0\ \right\rbr\ =\
-m\clr{{\ti T}_{01}}^{m},\label{csv4} 
\ea
once the conformal anomaly given by Eq. (\ref{tfanmc}) is  
put to zero.
%
                                                                                                                             
While Eqs. (\ref{gil1}) and (\ref{gil1})
carry the consequence of explicitly employing the Bessel-Hagen form of Eq. (\ref{cnc})
in two dimensions,
it turns out that the Fourier modes of the light-cone component of improved \et\
can be identified with the conformal generators
at the instant $\tau=0$.
To be concrete,
regard that
\ba
{\ti T}_{++} &=& 
\frac{1}{2}\left( {\ti T}_{00}+{\ti T}_{01}\right) -
\frac{1}{4} \left\lbc ({\ti T}_{01} - \ti T_{10})
+ ({\ti T}_{00} - {\ti T}_{11})\right\rbc, \\
{\ti T}_{--} &=& \frac{1}{2}\left( {\ti T}_{00}-{\ti T}_{01} \right)+
 \frac{1}{4} \left\lbc ( {\ti T}_{01} - {\ti T}_{10})
- ({\ti T}_{00} - {\ti T}_{11})\right\rbc. \ea
Meanwhile, the light-cone
components of the improved \et\
deviate from $\frac{1}{2}\left( {\ti T}_{00}+{\ti T}_{01}\right)$
and  $\frac{1}{2}\left( {\ti T}_{00}-{\ti T}_{01} \right)$
only by the anomalies. 
Hence, the light-cone derivatives of these light-cone 
components of the improved \et\ appear to be 
\ba
\left\lbr\ ({\ti T}_{00})^0- ({\ti T}_{01})^0,\ ({\ti T}_{++})^h\ \right 
\rbr&=&
\left\lbr\ ({\ti T}_{00})^0- ({\ti T}_{01})^0,\ \frac{1}{2}\left( 
{\ti T}_{00}+{\ti T}_{01}\right)^h \ \right \rbr\ =\ 0,\label{ccit1}\\
\left\lbr\ ({\ti T}_{00})^0+ ({\ti T}_{01})^0,\ ({\ti T}_{--})^h\ \right \rbr&=&
\left\lbr\ ({\ti T}_{00})^0+ ({\ti T}_{01})^0,\ \frac{1}{2}\left( 
{\ti T}_{00}-{\ti T}_{01}\right)^h \ \right \rbr\ =\ 0, \label{ccit2}
\ea
if the secondary constraints
induced by the anomaly-free conditions
given in Eqs. (\ref{tfanmc}) are also invoked.
Indeed, these respond to the conservation relations
\ba
\pt_{-}{\ti T}_{++} &=& \pt_{\bar z}T(z)\ =\ 0,\label{hl}\\
\pt_{+}{\ti T}_{--} &=& \pt_{z}{\bar T}({\bar z})\ =\ 0\label{ahl},
\ea
where Wick rotations are considered 
in accordance with Eqs. (\ref{z1}) and (\ref{z2}).
Therefore, the identifications of the Fourier modes of the 
improved \et\ with the generators of conformal transformations
in Eqs. (\ref{gil1}) and (\ref{gil2}) can be explicitly given as
\ba
e^{2m\tau}\clr{{\ti T}_{++}}^m&=&
\frac{1}{2}\int_{0}^{\pi} d\sigma e^{2m(\sigma+\tau)}
{\ti T}_{++}\ =\ \int_{0}^{2\pi} dx^1 e^{m(x^1+ix^0)} T(z)\ \equiv\ L_{m},\\ 
e^{-2m\tau}\clr{{\ti T}_{--}}^m&=&\frac{1}{2}\int_{0}^{\pi} d\sigma 
e^{2m(\sigma-\tau)}
{\ti T}_{--}\ =\ 
\int_{0}^{2\pi} dx^1 e^{m(x^1-ix^0)}{\bar T}(\bar {z})\ \equiv\ 
{\bar L}_{m},
\ea
where the variable $\sigma$ has been re-scaled in order to have 
the conventional range of integration.

The algebras of interest amount to the explicit  
derivation, on shell of Eq. (\ref{tfanmc}), as follows, 
\ba
\left\lbr\ \clr{ \frac{{\ti T}_{01}\pm {\ti T}_{00}}{2} }^m, 
\clr{ \frac{{\ti T}_{01}\pm {\ti T}_{00}}{2} }^n\ \right\rbr
&=& (n-m) \clr{ \frac{{\ti T}_{01}\pm {\ti T}_{00}}{2} }^{n+m} + 
\frac{D}{12}\clr{\frac{(m^3-m)-(n^3-n)}{2}}\delta_{m+n,0}\nr\\
&+& 
\clr{ \frac{ m(1-\delta_{n,0})-n(1-\delta_{m,0})}{16\pi^2 T} 
+\frac{(m-n)\delta_{n+m,0}}{8\pi^2 T}
} \int_{n+m}\clr {\frac{F\pm{\bar F}}{2}} \nr\\
&+&\frac{m(1-\delta_{n,0})-n(1-\delta_{m,0})}{16\pi^2 T}
\int_{n+m} \clr { \frac{{\bar g}_{\bar \ap}\pm g_{\ap}}{2} } \nr\\
&-&  \frac{ (m^3-m)(1-\delta_{n,0})-(n^3-n)(1-\delta_{m,0})} {12\pi^3 T^2} 
\int_{n+m}tr \left( G\clr{\frac{ {\cal E}_{2} \pm  {\bar {\cal E}}_{2} }{2}}\right)
\nr\\
&-& \clr{ \frac{ m(1-\delta_{n,0})-n(1-\delta_{m,0}) 
}{48\pi^3 T^2}
-\frac{(m-n)\delta_{n+m,0}}{24\pi^3 T^2}
-\frac{(m-n)mn}{24\pi^3 T^2} } \nr\\
&\cdot & \int_{n+m}tr \clr {G \clr{
\frac{{\bar {\cal E}}_{1}\pm {\cal E}_{1}}{2}
+\frac{{\bar {\cal E}}_{3}\pm {\cal E}_{3}}{2}
} }\nr\\ 
&-& \frac{nm(m-n)}{16\pi^3 T^2}\int _{n+m}\frac{DY \pm DY}{2}
+O(5)+\cdots,\label{galg}
\ea
where
\ba
g_{\ap} &=& 
\frac{1}{T^2}\Pi G(1-\ap_{3}){\cal E}_{2}G\Pi+
\ac X(1-\ap_{4}){\cal E}_{2}\ac X +\frac{2}{T}\Pi G \clr{(1-\ap_{1})
{\cal E}_{1} +(1-\ap_{2}){\cal E}_{3}} \ac X,\label{gap1}\\
{\bar g}_{\bar \ap} &=&
\frac{1}{T^2}\Pi G(1-{\bar \ap}_{3}){\bar {\cal E}}_{2}G\Pi+
\ac X(1-{\bar \ap}_{4}){\bar {\cal E}}_{2}\ac X +\frac{2}{T}\Pi G \clr{(1-
{\bar \ap_{1}}){\bar {\cal E}}_{1}+ 
(1-{\bar \ap}_{2}){\bar {\cal E}}_{3}} \ac X.\label{gap2} 
\ea
Remarkably, the dependence of algebras
of generators on the improvement terms
is observed in Eqs. (\ref{galg}).
Thus, the closure of algebras for 
the light-components of \et\
could not be a generic feature.
In addition,
both the two algebras mentioned above 
can not be simultaneously closed up to c-numbers 
in view of the particular structure given in terms of the $\pm$ signs.
Therefore, it may be of interest 
whether there exist some particular choices of \fv s for which one of the 
algebras of Eqs. (\ref{galg}) may reproduce the
desired closed Virasoro version.
This situation seems to be realised by
requiring either ${\cal E}_{i}$'s $={\bar {\cal E}}_{i}$'s 
or ${\cal E}_{i}$'s $=-{\bar {\cal E}}_{i}$'s, 
each of which is accompanied with the conditions $\ap_{i}$'s $={\bar \ap}_{i}$'s, 
but none of these solutions are permissible to the 
consistency equations given in Eqs. (\ref{con1} - \ref{con8}).
However, the possibility for
further compacting the form of Eqs. (\ref{galg}) 
remains.
For example, considerable simplifications can be achieved by means of 
putting the quantity $DY$ zero and/or discarding some of the 
traceless parts of the six matrices for the improvement terms.
The traceless part of matrices  
are reducible because
they are restrained only by four of the consistency conditions
in Eqs. (\ref{con1} - \ref{con8}).

For consistency check, 
it is  worthwhile to note that all the coefficients 
for the central extension in Eqs. (\ref{galg})
tend to vanish when one of the 
generators are given in the zeroth mode. 
The corresponding structure may also be formally
derived from Eqs. (\ref{galg}), namely,
\ba
\left\lbr\ \clr{\frac{{\ti T}_{01}\pm {\ti T}_{00}}{2}}^0,
\clr{\frac{{\ti T}_{01}\pm {\ti T}_{00}}{2}}^n\ \right\rbr
&=& 
\left\lbr\ \clr{\frac{{\ti T}_{01}}{2}}^0, \clr{\frac{{\ti T}_{01}\pm 
{\ti T}_{00}}{2}}^n\ \right\rbr
\pm \left\lbr \clr{\frac{{\ti T}_{00}}{2}}^0,
\clr{\frac{{\ti T}_{01}\pm {\ti T}_{00}}{2}}^n\ 
\right\rbr\nr\\
&=& 
\left\lbr\ \clr{\frac{T_{01}}{2}}^0, \clr{\frac{{\ti T}_{01}\pm 
{\ti T}_{00}}{2}}^n\ \right\rbr
\pm \left\lbr \clr{\frac{T_{00}}{2}}^0,
\clr{\frac{{\ti T}_{01}\pm {\ti T}_{00}}{2}}^n\
\right\rbr\nr\\
&=&
n\clr{ \frac{{\ti T}_{01}\pm {\ti T}_{00}}{2}}^n,
\ea
based on Eqs. (\ref{csv1} -\ref{csv4}).
Certainly, the conservation relations above 
coincides with the ones suggested by the Virasoro version.

\section{Discussions}
\label{five}
To summarise, the quantisation should not be restricted by the field content
prescribed a priori;
field contents themselves pertains to the properties of the quantised theory itself.
The incorporation with \fr\ exploits the possibility to end up the quantisation
with permissible configurations against the conventional starting point.

Considerably,
the method of canonical quantisation
further arms the traditional field theoretical treatments
with extra interesting information.
The corresponding intuitive implementation for the \fr\
appears more viable, in contrast to those within the path integral approaches, 
by means of its relevance to the 
canonical transformation among dynamically equivalent systems \cite{jw,tm}, even though 
further generalisation remains possible \cite{cnn}. 
The ambiguity contribution for various operators and their algebras
thus responds to reformulating the theory with alternative \fv s.
Specifically,
this leaves freedom appropriate for
the historical Schwinger terms \cite{tgi,swg}
in the current algebra.
Moreover,
as it was shown in section \ref{two}, 
nor can be the form for the charge algebra 
heuristically predicted by its classical version
since the quantisation entails the \fr\ that 
exceeds beyond the scope of the appending total divergence to the current operators.
                                                                                                                             
In practise, speaking of the results for the nonlinear sigma model
in Section \ref{three}, 
the \fr\ in the Hamiltonian scheme first retains the embedding
flexibility for quantisation and then regains the consistency 
by deviating field properties 
from the convention scalar prescription after a simple covariant requirement \cite{jw}.
Noteworthy is only that
the use of the term {\et stress energy tensor} is prevented
in our formulation.
Derived according to the variation with respect to
the space-time metric, the meaning of such term
is less concrete in the Hamiltonian dynamics of
flat space-time, even though possible linkages of this to the Belifante
tensor could be imagined \cite{jk1,ccj}.
Furthermore, the conformal anomaly conditions indeed
appear to be less strict than the conventional Weyl anomaly conditions
\cite{tm,cf,cf1,ts,mt,cl}.
Remarkably, the additional restrictions, e.g. $D=26$,
are originated from the reparametrisation invariance
that imposes the equivalence between the Weyl and conformal symmetry
on the two dimensional field theories \cite{sj,cl}.
                                                                                                                             
The calculation in Section \ref{four}
explicitly probes the quantised algebras
at the charge level.
We find it impossible to recover the
the closed conformal algebras, i.e., the Virasoro version,
at the two loop level for the problem at hand.
However, 
the conformal currents conserve whenever the symmetry is recovered.
Subsequently, the compatibility
with the conventional closure expectation based on the earlier lower-order treatments \cite{tb}
remains undisturbed.
This is because the discrepancy for the conformal algebra is
only up to $O(4)$ which corresponds to conventional two-loop level,
where the symmetric and traceless properties
for the canonical \et\ have been lost.
                                                                                                                             
To conclude, 
exploiting the formulation
in terms of alternative sets of \fv s should of fundamental interests for any models.
While the \fr\ responds to generalising the field content, 
the present study the nonlinear sigma model in fact offers consistent issues:
In the historical treatment of Yang-Mill models \cite{jk1,dw,jka},
it was known that the gauge invariant conformal charge 
cannot form the closed algebras. 
This does not actually bother the conformal symmetry 
but exhibits the non-integrable structure \cite{jka,yang}
in response to the path-dependent finite transformation of gauge fields.
Hence, vector potential could be important ingredient for the conformal symmetry
equipped with algebra of nontrivial form.
On the other hand, 
the \fv s of nontrivial spin structure
may enter \cite{jk1} into the nonlinear sigma model, 
and respond to the vector fields, 
by means of the failure of canonical \et\
to be manifestly symmetric and traceless at the two loop order.
Moreover, gauge potentials are essential components 
for the study of the fermionization of 
nonlinear sigma model\cite{pw,wi};  
the presence of the such degree of freedom need not come as surprise.
Our work attempts to address itself to the concrete picture 
and to the further applicability for the canonical quantisation.
Of course, more open questions are left,
in a straightforward manner.
These may concern formulating on the curved space-time, 
explicit incorporating gauge fields or even the super-fields.
However, we close the present work
by further claiming the following fundamental point.
The employment of the Hamiltonian \fr\ subject to preservation
of canonical commutators needs not be the most generic handling 
of field theory \cite{cnn}.
We hope that through all the efforts made on the nonlinear sigma model
more insight for the field
theory in general can be revealed.

\section{acknowledgement}
The authors would like to thank Wolfdieter Lang for 
his thorough reading and valuable suggestions for this article.
Part of the work has been inspired by Stefan Theisen.

\end{document}